\documentstyle[sprocl]{article}

\def\Journal#1#2#3#4{{#1} {\bf #2}, #3 (#4)}

\def\ApJ{\em Astrophys. J.}
\def\ApJL{\em Astrophys. J. Lett.}
\def\AJ{\em Astron. J.}

\def\be{\begin{equation}}
\def\ee{\end{equation}}
\def\bea{\begin{eqnarray}}
\def\eea{\end{eqnarray}}

\begin{document}

\title{BEYOND HDF - SEARCHING FOR EARLY STAR FORMATION IN THE INFRARED}
\author{ D. THOMPSON }
\address{Max-Planck-Institut f\"ur Astronomie\\
K\"onigstuhl 17\\
D-69117 Heidelberg, Germany}
\author{}
\address{}

\maketitle\abstracts{
The success of the Hubble Deep Field (HDF) data in identifying galaxies at
redshifts up to $\sim$3 has been quite spectacular.  It is possible to extend
this to even higher redshifts using infrared techniques, several of which 
are briefly described in this paper.}

\section{Introduction}
Other papers in this volume describe some of the results which have come 
from the HDF data.  I describe here several types of observations in the 
infrared which can take us beyond the redshifts probed by the HDF.
Although I have split this paper into two main sections, one covering
emission-line techniques and the other covering continuum-based techniques,
it should be noted that all of these survey methods make use of some 
strong spectral feature in order to select galaxies out of a background of
field galaxies at lower redshift.

\section{Emission-Line Surveys}


Stretching the intended meaning of {\em infrared} to include CCD-based 
narrowband surveys at wavelengths beyond 7000\,\AA\ includes searches
for Ly$\alpha$-bright galaxies up to a redshift approaching 7.  Field 
surveys for such objects\,\cite{none} have, to date, been unsuccessful 
at identifying any significant population of galaxies, though observations
targeting existing structures,\cite{none} such as quasar absorption-line
systems of known redshift, have identified a number of interesting 
objects.  Because the Ly$\alpha$ line is resonant, and can suffer 
multiple scattering off atomic hydrogen as it passes through the ISM, 
the chances of absorption by dust grains is proportionally higher.  This 
dust {\em quenching} of the Ly$\alpha$ line is generally thought to explain 
the lack of success in these field surveys.

Recent, detailed models by Thommes \& Meisenheimer,\cite{th} including 
both dust formation and a scatter in the time when massive star formation 
begins, give considerably more pessimistic predictions on the volume density 
of forming galaxies than the canonical results of Baron \& White.\cite{bw}  
Even so, it should still be possible to identify high-redshift galaxies by
targetting their Ly$\alpha$ emission redshifted into the CCD infrared.
The Calar Alto Deep Imaging Survey\,\cite{km} (CADIS) is attempting to 
cover sufficient volume and depth to reach these new limits, and several 
$z > 5$ candidate objects have been identified to date.



If sufficient dust is generated early on in a starburst to completely 
destroy the Ly$\alpha$ photons, then it would still be possible to 
detect the starbursts through other emission lines, most notably the
restframe optical lines of [O II]\ 3727\,\AA, H$\beta$\ 4861\,\AA, 
[O III]\ 5007\AA, and H$\alpha$\ 6563\,\AA.  At high redshift, these lines
are shifted into the near infrared $JHK$ bands, where they can be imaged
through narrowband filters.

This technique has only been practical since the development of
reasonably large infrared arrays.\cite{tdb}$^,$\cite{mb}  There are 
several groups currently engaged in surveys using this
technique,\cite{none}$^,$\cite{tmb}$^,$\cite{mtm} with a number of 
objects having already been identified.  These surveys primarily target 
the H$\alpha$ line at $z \simeq 2.4$, where it is redshifted into the 
$K$ band, but the method is sensitive to any emission lines, and 
can be used to image [O II]\ 3727\,\AA\ at redshifts as high as five.

\section{Continuum Surveys}

Complementary to the emission-line surveys are those based on continuum
features.  Steidel et al.\cite{st} have used the Lyman limit feature at
a restframe wavelength of 912 \AA\ with remarkable success to identify 
a population of galaxies at $z \simeq 3.25$, many without strong emission
lines despite relatively high star formation rates.

If we try to push this technique into the near infrared, we run into 
problems with the continuum depression across the Ly$\alpha$ line, which 
increases dramatically at redshifts greater than four.\cite{ken}
Extrapolating to $z \simeq 7$, which places the Ly$\alpha$ line between the
$I$ and $J$ bands, one would expect that virtually all of the flux blueward of
the Ly$\alpha$ line would be absorbed.  These $I$-band drop-out objects would 
appear around $J \simeq 23.4$ (Johnson), assuming little extinction from 
dust, for star formation rates of $\simeq 50$\,M$_\odot$\,yr$^{-1}$.  Such
limits are technologically feasible with the current generation of infrared
arrays on 4m-class telescopes.


Rather than searching directly for forming galaxies at high redshift, 
another method of probing the earliest epoch of galaxy formation is to 
look for old, evolved objects at lower redshift.  Evolved, or passively
evolving, stellar populations, such as found locally in elliptical galaxies,
can develop a strong break in their spectra around 4000\,\AA\ a few hundred
million years after their last burst of star formation, giving another 
spectral feature on which to base a survey.  Sufficiently evolved objects at
$z > 1$ can imply formation redshifts of $z > 3$. 

As this 4000\,\AA-break feature is redshifted through the optical bands, 
the optical-to-infrared colors of these objects becomes increasingly red.
At $z \simeq 1.5$, this break lies between the $I$ and $J$ bands.  Deep, 
multicolor imaging in the near infrared can thus distinguish these ``extremely
red objects'' from foreground galaxies.  This is potentially a new population 
of objects; while relatively easy to detect at near infrared wavelengths, 
no significant numbers of these objects would have been included in even the
deepest, optically-selected redshift surveys.  Several such galaxies have
recently been identified\,\cite{dun}$^,$\cite{gd} from serendipitous 
observations, though the true extent of any field population is largely 
unknown.  A large-scale field survey would thus be valuable, to determine
the space density of such objects and produce a sample for further study.

\section{Conclusions}

The development of large infrared arrays has opened up several possibilities 
for pushing beyond the redshifts probed by the HDF data.  Current 
ground-based efforts prehaps presage what we might expect from the NICMOS
camera on the Hubble Space Telescope.

\end{document}